# RESET THRESHOLDS OF TRANSFORMATION MONOIDS

Igor Rystsov
National Technical University of Ukraine, Kiev, Ukraine

Marek Szykuła
University of Wroclaw, Faculty of Mathematics and Computer Science, Wroclaw, Poland

**Abstract.** Motivated by the Černý conjecture for automata, we introduce the concept of monoidal automata, which allows the formulation of the Černý conjecture for monoids. We show upper bounds on the reset threshold of monoids with certain properties. In particular, we obtain a quadratic upper bound if the transformation monoid contains a primitive group of permutations and a singular of maximal rank with only one point of contraction.

## 1. Introduction

We consider finite automata together with their transition monoids. A deterministic finite complete semiautomaton, called later *automaton*, is synchronizing if it admits a reset word, whose action sends all states to one state. The reset threshold of a synchronizing automaton is the minimum length of its reset words. Synchronizing automata are most famous due to the Černý conjecture from 1969, which says that the reset threshold of a synchronizing automaton with $n$ states is at most $(n-1)^2$ [1].

In this work, we study the problem from the perspective of monoid properties. We define the reset threshold and the Černý conjecture for finite transformation monoids, which are equivalent to that for ordinary finite automata but stated apart from a particular set of generators.

The concept is not entirely new. Similar ideas appeared in many papers but in an implicit way. For example, Cayley graphs were used for groups as early as the 19th century, and in [2] they are used to investigate the Černý conjecture. In [3], there are considered some varieties of monoids and, in fact, determined their reset thresholds. From a similar standpoint, the Černý conjecture was also considered in [4], where lower and upper bounds on the reset threshold of the full transformation monoid were obtained. In [5], a theory of transformation monoids using the theory of categories was developed.

## 2. Our contributions

First, the concept of the reset threshold for transformation monoids and monoidal semiautomata is introduced. Then we consider the class of transformation monoids whose digraph of singulars is strongly connected. Our main result is a quadratic upper bound on the reset threshold of a monoidal semiautomaton whose transformation monoid belongs to that class and if there is a generator that has only one point of contraction. In other words, these are transformations with the kernel type $(k, 1, \dots, 1)$, where $k \geq 2$. We also show that if $k = 2$, then such semiautomaton is completely reachable.

As a corollary, we obtain a quadratic upper bound on the reset threshold of monoids with a primitive group that also has a non-invertible transformation of the mentioned kernel type. The results also give a partial answer to Problem 12.36 of [6]. Finally, we obtain a tight bound on the reset threshold of a symmetric inverse monoid.

## 3. Preliminaries

Let $S$ be a finite set consisting of $n$ elements, where $n > 2$. The elements of $S$ are called *states* and we assume that $S = \{1, \cdots, n\}$. The full monoid of transformations on the set $S$ we denote by $\mathrm{End}_n(S) = \{f \mid f: S \longrightarrow S\}$. The elements of this monoid are called *maps* (or *self-maps*).

We assume that a map $f$ acts from the right and denote by $(s)f$ the image of the state $s$ under the action of $f$. The image of a map $f$ we denote by $\mathrm{im}(f) = \{(s)f : s \in S\}$ and its completion by $\mathrm{coim}(f) = S \backslash \mathrm{im}(f)$, which is called the *coimage* of $f$. The equivalence $\ker(f) = f \circ f^{-1}$ is the kernel of a map $f$. The number $|\mathrm{im}(f)|$ is called the rank of the map $f$ and is denoted by $\mathrm{rk}(f)$. The number $\mathrm{crk}(f) = |\mathrm{coim}(f)| = n - \mathrm{rk}(f)$ is called the corank of *f.*

Maps of corank zero are permutations (bijections) on the set $S$. Maps of nonzero corank are called singulars. Note that each singular merges at least one pair of states. A simple singular $f$ is a map of unit corank and its kernel type is $(2,1, \cdots, 1)$. Each simple singular merges exactly one pair of states. A one-point singular is a singular with one point of contraction whose kernel type is $(k, 1, \cdots, 1)$, where $k \geq 2$ [7]. Therefore, a simple singular is a special case of a one-point singular. A constant map is a map of unit rank. The set of all $n$ constants from the full monoid $\mathrm{End}_n(S)$ denoted by $\mathrm{End}_1(S)$.

**Definition 1.** A submonoid of the full monoid $\mathrm{End}_n(S)$ is called a *transformational monoid on* the set $S$ or simply a *monoid on* the set $S$.

Before turning to monoids on the set $S$, we have to consider monoid actions on the set $S$. Fix a monoid $M$, then a *transition function* $\varphi: S \times M \longrightarrow S$ is a function that satisfies the following equalities for all $s \in S$, and $x, y \in M$:

$$\varphi(s, xy) = \varphi(\varphi(s, x), y), \qquad \varphi(s, 1_M) = s, \tag{1}$$

where $1_M$ is the identity of $M$ [5].

Thus, the transition function $\varphi$ defines the action of monoid $M$ on the set $S$. For an element $m \in M$ we define the map $\varphi_m: S \longrightarrow S$ by the formula $(s)\varphi_m = \varphi(s, m)$ for all $s \in S$. From (1), we have $\varphi_{xy} = \varphi_x \circ \varphi_y$ and $\varphi_1 = i_S$, where $i_S$ is the identity function on $S$. Therefore, the action of monoid $M$ on the set $S$ induced the morphism (homomorphism) $\tilde{\varphi}: M \longrightarrow \mathrm{End}_n(S)$ of monoids, which is defined by the formula $\tilde{\varphi}(m) = \varphi_m$, for all $m \in M$.

If the morphism $\tilde{\varphi}$ is injective, then the monoid $M$ is isomorphic to the monoid $\tilde{\varphi}(M)$ on the set $S$. In the remaining part of the paper, we consider only this case. A monoid $M$ on the set $S$ defines the action $\varphi(s, f) = (s)f$ for all $s \in S$, and $f \in M$. In this case, the transition function $\varphi$ reduced to applying a map to a state and so, it can be omitted.

**Definition 2.** A monoid $M$ on the set $S$ is called *transitive* if for every two states $s, t \in S$ there exists a map $f \in M$ such that $(s)f = t$. A monoid $M$ on the set $S$ is called *synchronizing* if $M$ contains a constant map.

Note that a constant map moves all states to the same state. The following characteristic for synchronizing automata is well known [8] and can be easily proved for monoids.

**Proposition 1.** A monoid $M$ on the set $S$ is synchronizing if and only if for every pair of states $s, t \in S$, there is a map $f \in M$ such that $(s)f = (t)f$.

## 4. Automata

Let $X \subseteq \mathrm{End}_n(S)$, then denote by $\langle X \rangle$ the monoid on $S$ generated by the set of maps $X$.

**Definition 3.** A *monoidal semiautomaton* $A$ (*M-automaton* or simply *automaton*) is a pair $A = (S, X)$, where $S$ is a set of states and $X \subseteq End_n(S)$.

The monoid $\langle X \rangle$ on the set $S$ is called the *transition monoid* of an automaton $A = (S, X)$. In the sequel, we assume that an automaton $(S, X)$ inherits all the properties of its transition monoid $\langle X \rangle$, such as the transitivity, the synchronizability, and so on. When the context is clear, we also call a monoidal semiautomaton shortly an automaton.

Note that a usual finite semiautomaton $A = (S, \Sigma, \delta)$ can be treated as the monoidal semiautomaton $(S, X)$, where $X = \{\delta_a : a \in \Sigma\}$ is the set of its basic maps $\delta_a(s) = \delta(s, a)$ for all $s \in S$. In this case, the monoid $\langle X \rangle$ is equal to the transition monoid of $A$. The difference between $A$ and the monoidal semiautomaton is the explicit existence of $\Sigma$ of the former, which allows $A$ having e.g., two equal maps labeled by different letters, whereas $X$ is not a multiset. Such differences are, however, irrelevant to our problem.

A set of generators $X$ allows to define the measure of complexity for a map decomposition on generators, namely, let the length $l_X(f)$ of a map $f \in M$ with respect to $X$ be the smallest length of a representation $f$ as a composition (product) of maps from $X$. The introduced measure (metric) has the following property, which is similar to the logarithm property:

$$l_X(f \cdot g) \leq l_X(f) + l_X(g) \ , \qquad (2)$$

for all $f, g \in M$.

Let $A = (S, X)$ be a synchronizing automaton, then the reset threshold $\mathrm{rt}(A)$ of $A$ is the minimum length of its constants:

$$\mathrm{rt}(A) = \min_c \{\ l_X(c) : c \in \langle X \rangle \cap \mathrm{End}_1(S)\ \} \,.$$

The well-known Černý conjecture [3] can be formulated for semiautomata in the following way.

**Conjecture 1.** For a synchronizing monoidal semiautomaton $A = (S, X)$ with $n$ states, we have $\mathrm{rt}(A) \le (n-1)^2$.

If $A = (S, X)$ and $B = (S, Y)$ are synchronizing automata and $Y \subseteq X$, then the following inequality holds:

$$\mathrm{rt}(A) \le \mathrm{rt}(B) \,, \tag{3}$$

since any decomposition of a constant in the automaton $B$ is also its decomposition in the automaton $A$.

The *reset threshold* of a synchronizing monoid $M \le \mathrm{End}_n(S)$ is the maximum reset threshold of automata, taken over all sets of its generators $X \subseteq M$:

$$\mathrm{rt}(M) = \{\mathrm{rt}(A) : \ A = (S, X),\ M = \langle X \rangle\} \ . \tag{4}$$

Thus, Černý conjecture can be formulated for monoids in the following equivalent way.

**Conjecture 2.** For every synchronizing monoid $M \le \mathrm{End}_n(S)$ on the set $S$, the following inequality holds: $\mathrm{rt}(M) \le (n-1)^2$.

Thus, the Černý conjecture cannot be formulated without monoid generators, but if it is true, then it does not depend on generators.

## 5. Directable automata

Let $M$ be a monoid on the set $S$ and $T \subseteq S$ be a subset of states. Then a map $f \in M$ is called *augmenting for* $T$ if $|(T)f^{-1}| > |T|$. A monoid $M$ on the set $S$ is called *directable* if $M$ contains augmenting maps for all proper subsets of $S$ (nonempty and unequal to $S$). The following characteristic is well known for finite automata [8], which we prove for completeness.

**Lemma 2.** A monoid $M$ on the set $S$ is directable if and only if it is synchronizing and transitive.

**Proof.** Let $M$ be a directable monoid on the set $S$, then fix an arbitrary state $s_1 \in S$ and successively augment the subset $\{s_1\}$ with augmenting maps $f_1, \cdots f_k$ until the entire set $S$ is obtained. Then the map $c = f_k \cdot \cdots \cdot f_1$ will be a constant since $S = (s_1)c^{-1}$ and hence $\mathrm{im}(c) = \{s_1\}$. Thus, $M$ is synchronizing and it remains to note that $(s)c = s_1$ for any $s \in S$. Because a state $s_1$ can be chosen arbitrarily, it follows that $M$ is transitive.

Conversely, let $M$ be a synchronizing and transitive monoid on the set $S$ and let $T \subset S$ be a nonempty subset of states. Take a constant $c_1 \in M$, which maps all states to a state $s_1$, and take map $f \in M$ such that $(s_1)f \in T$. Then for the map $c_2 = c_1 \cdot f$ we have $(T)c_2^{-1} = S$. Since $|T| < |S|$ the map $c_2$ will be augmenting for $T$ and the lemma is proved. □

Recall that an automaton $(S, X)$ is directable if the monoid $\langle X \rangle$ is directable. Let $A = (S, X)$ be a directable automaton and $T$ be a proper subset of $S$, then we denote by $\mathrm{at}_A(T)$ the minimum length of a map that augments $T$. The *augment threshold* $\mathrm{at}(A)$ of a directable automaton $A = (S, X)$ is defined by the formula:

$$\mathrm{at}(A) = (\mathrm{at}_A(T) : \emptyset \subset T \subset S) \ . \qquad (5)$$

The following lemma is also well known for finite automata [8].

**Lemma 3.** Let $A = (S, X)$ be a directed automaton with $n$ states, then the following inequality holds $rt(A) \leq at(A) \cdot (n-2) + 1$.

**Proof.** Proposition 1 implies that there is a singular $f_1 \in X$, which merges at least one pair of states. Then we can choose a state $s_1$ such that $|(s_1)f_1^{-1}| > 1$. Then we can successively augment the subset $(s_1)f_1^{-1}$ by augmenting maps $f_2, \cdots, f_k$ such that $l_X(f_i) \leq at(A)$, $2 \leq i \leq k$, until the entire set $S$ is obtained. Therefore, we have decomposition $c = f_k \cdot \cdots \cdot f_1$ of a constant $c$ as in Lemma 2. Note that $k \leq n - 1$, and we obtain from the construction and property (2) the inequalities:

$$l_X(c) \leq \mathrm{at}(A) \cdot (k-1) + 1 \leq \mathrm{at}(A) \cdot (n-2) + 1 \ .$$

Thus, the lemma is proved. □

## 6. One-point automata

Consider the properties of a one-point singular $f$ of the kernel type $(k, 1, \cdots, 1)$ in more detail. There is a unique state $d_f \in \mathrm{im}(f)$, whose preimage consists of $k$ states (*duplicate state*). There is also a unique state $c_f \in (f^{-1}(d_f) \cap \mathrm{im}(f))$ that is in the same $f$-cycle as the state $d_f$. Note that the states $c_f$ and $d_f$ may coincide, in this case, the state $d_f$ is in a loop $(d_f)f = d_f$. The coimage $S \backslash \mathrm{im}(f)$ consists of $k-1$ states, which we denote by $E_f = \{e_1, \cdots, e_{k-1}\}$ (excluded states), or one state $\{e_f\}$ if $k = 2$. We denote all other states by $I_f = \mathrm{im}(f) \backslash \{d_f\}$. Thus, a one-point singular $f$ has the following properties:

$$|(s)f^{-1}| = 0,\ s \in E_f,\quad \left|\left(d_f\right)f^{-1}\right| = k,\quad |(s)f^{-1}| = 1,\ s \in I_f\ . \qquad (6)$$

**Lemma 4.** If $d_f \in T$ and $\mathrm{coim}(f) \nsubseteq T$, then the one-point singular $f$ augments the subset $T$.

**Proof.** Since the preimages of states do not intersect and $\left|E_f \cap T\right| \leq k-2$, we obtain from property (6) the following inequality:

$$|(T)f^{-1}| - |T| = |(T)f^{-1}| - \left|E_f \cap T\right| - \left|\{d_f\} \cap T\right| - |I_f \cap T| =$$

$$= \sum_{s \in E_f \cap T} |(s)f^{-1}| - \left|E_f \cap T\right| + \left|\left(d_f\right)f^{-1}\right| - 1 + \sum_{s \in I_f \cap T} |(s)f^{-1}| - |I_f \cap T| \geq$$

$\geq\ -(k-2) + (k-1) + 0 = 1$ . □

Let $\mathrm{Aut}_n(S) = \{g \mid g: S \leftrightarrow S\}$ be the symmetric group of all permutations on the set $S$. Note that the group $Aut_n(S)$ is a subgroup of invertible maps of the full monoid $\mathrm{End}_n(S)$. A subgroup of the symmetric group $G \leq \mathrm{Aut}_n(S)$ is called a permutation group on the set $S$ or simply a group on the set $S$.

A group $G$ on the set $S$ defines the action $\varphi(s,g) = (s)g$ on $S$ and strongly connected components of this action are called the *orbits* of $G$. The *orbit of a state* $s \in S$ is equal to $(s)G = \{(s)g : g \in G\}$. A group $G$ on the set $S$ is called transitive if for every two states $s, t \in S$ there exists a permutation $g \in G$ such that $(s)g = t$. In this case, the group $G$ has only one orbit, which equals $S$.

The action of a group $G$ on the set $S$ extends to the square $S \times S$ by components $(s,t)g = ((s)g, (t)g)$ for all $g \in G$ and $s, t \in S$. For a relation $\rho \subseteq (S \times S) \backslash \Delta_S$ and $H \subseteq G$ we put $(\rho)H = \{(s,t)g : (s,t) \in \rho, g \in H\}$. The orbits of a group $G$ on the set $(S \times S) \backslash \Delta_S$ are called the *orbitals* of this group. The *orbital* of a pair $(s,t) \in (S \times S) \backslash \Delta_S$ is equal to $(s,t)G = \{(s,t)g : g \in G\}$. The group closure of a relation $\rho \subseteq (S \times S) \backslash \Delta_S$ is the union of orbitals:

$$(\rho)G = \bigcup_{(s,t)\in\rho} (s,t)G \,.$$

**Definition 4.** An automaton $(S, X)$ is called *one-point* if $X = \{f\} \cup Y$, where $f$ is a one-point singular and $Y = \{g \in X : crk(g) = 0\}$ are permutations on the set $S$. If $f$ is a simple singular then we call one-point automata $A = (S, \{f\} \cup Y)$ as *weakly singular* automata.

Let $A = (S, \{f\} \cup Y)$ be a one-point automaton and let $G = \langle Y \rangle$ be its permutation group, then we put $\pi_0 = E_f \times \{d_f\} = \{(e_i, d_f) : 1 \le i \le k-1\}$. We define the relations $\pi_m$ for $m \ge 1$ by the formula:

$$\pi_m = \{(e_i, d_f)g : e_i \in E_f,\ g \in G,\ l_Y(g) \le m\}. \quad (7)$$

It is easy to see that $\pi_m = (\pi_{m-1})Y \cup \pi_{m-1}$, so relations $\pi_m$ form an increasing chain, which ends by the group closure relation $(\pi_0)G$:

$$\pi_0 \subset \cdots \subset \pi_m \subset \cdots \subset (\pi_0)G \,. \quad (8)$$

We associate the digraph $\Gamma(\rho) = (S, \rho)$ with a relation $\rho \subseteq (S \times S)\backslash\Delta_S$, where $S$ is the set of its vertices and $\rho$ is the set of its arcs. A binary relation $\rho$ is called strongly connected (or transitively complete) if its transitive closure $\rho^*$ is equal to $S \times S$. In this case, the digraph $\Gamma(\rho)$ will be strongly connected.

**Lemma 5.** If for a one-point automaton $A = (S, \{f\} \cup Y)$ the relation $\pi_m$ in the chain (8) is strongly connected then for every proper subset of $S$ there is an augmenting map of the form $f \cdot g$, where $g \in G$ and $l_Y(g) \le m$.

**Proof.** By definition, the digraph $\Gamma(\pi_m)$ is strongly connected. Let $T \subset S$ be a proper subset of states, then there is an arc $((e_i)g, (d_f)g)$ of $\Gamma(\pi_m)$ such that $(e_i)g \notin T$ and $(d_f)g \in T$, since otherwise, it would be impossible to exit the subset $S\backslash T$ along the arcs of $\Gamma(\pi_m)$. From this, it follows that $e_i \notin (T)g^{-1}$ and $d_f \in (T)g^{-1}$, so $\mathrm{coim}(f) \nsubseteq (T)g^{-1}$ and $d_f \in (T)g^{-1}$. Then according to Lemma 4, the one-point singular $f$ augments the subset $(T)g^{-1}$. In this case, the map $f \cdot g$ augments $T$, since the conditions satisfy $(T)(f \cdot g)^{-1} = ((T)g^{-1})f^{-1}$ and $|T| = |(T)g^{-1}|$. Thus, the lemma is proved.□

We denote the relation $(\pi_0)G$ by $\pi_A$ and call it the *singular relation* of a one-point automaton $A$. From Lemma 5, we have the following statement.

**Corollary 6.** If for a one-point automaton $A = (S, \{f\} \cup Y)$ the relation $\pi_A$ is strongly connected then for every proper subset of states $T$ there is an augmenting map of the form $f \cdot g$, where $g \in G$.

Let $M$ be a monoid on the set $S$, then denote by $\mathrm{im}(M)$ the set of all its images $\mathrm{im}(M) = \{\mathrm{im}(f) : f \in M\}$. A subset $T \subseteq S$ is called *reachable* in an automaton $(S, X)$ if $T \in \mathrm{im}(\langle X \rangle)$. Denote by $\mathrm{Sub}(S)$ the set of all nonempty subsets of $S$. An automaton $(S, X)$ is called *completely reachable* if $\mathrm{im}(\langle X \rangle) = \mathrm{Sub}(S)$ [9].

The following theorem can be also inferred from the characterization [10], which is, however, more complicated.

**Theorem 7.** Let $A = (S, \{f\} \cup Y)$ be a weakly singular automaton with the strongly connected relation $\pi_A$, then $A$ is completely reachable.

**Proof.** Let us show by induction on $n - |T|$ that every nonempty subset $T$ is reachable. For identity map $i_S \in \langle X \rangle$, we have $im(i_S) = S$, so the set $S$ is reachable. Suppose by induction that all subsets containing $m + 1$ states are reachable and let $|T| = m$. From corollary 6 it follows that for $T$ there is an augmenting map $f \cdot g$, where $g \in \langle Y \rangle$. Let us put $U = ((T)g^{-1})f^{-1}$, then $|(T)g^{-1}| = m$ and therefore $|U| = m + 1$, since $f$ is a simple singular. According to the assumption of induction, the subset $U$ is reachable, thus there is a map $h$ such that $\mathrm{im}(h) = U$. Further, we have $(U)f \subseteq (T)g^{-1}$, and $|(U)f| \geq m = |(T)g^{-1}|$. Hence $(U)f = (T)g^{-1}$ and therefore $(U)fg = (T)g^{-1}g = T$. From this, we conclude that $\mathrm{im}(hfg) = T$ and the theorem is proved by induction. □

Hence, a quadratic upper bound on the reset threshold of a weakly singular automaton with the strongly connected relation $\pi_A$ follows from [11]. Yet, we are going to prove it for a wider class of one-point automata, which are not necessarily completely reachable.

Let $A = (S, \{f\} \cup Y)$ be a one-point automaton and let us assume that the relation $\pi_A$ is strongly connected, then we denote by $\mathrm{msc}(A)$ the smallest number $m$, from which all relations in the chain (8) are strongly connected. Then from Lemma 5 follows that an automaton $A$ will be directional and the inequality holds:

$$\mathrm{at}(A) \leq \mathrm{msc}(A) + 1\,, \tag{9}$$

where $\mathrm{at}(A)$ is the augment threshold of $A$, which was defined by formula (5).

A pair $(s,t) \in \rho$ and an arc of the digraph $\Gamma(\rho)$ are called *cyclic* if the arc $(s,t)$ located in the cycle of the digraph $\Gamma(\rho)$. A relation $\rho \subseteq (S \times S)\backslash\Delta_S$ and the digraph $\Gamma(\rho)$ are called cyclic if every its pair (arc) is cyclic. In this case, the digraph $\Gamma(\rho)$ will be a union of cycles and disjoint union (by vertices) of strongly connected digraphs [12]. The cyclic part $C(\rho)$ of $\rho$ is the union of all its cyclic pairs.

**Lemma 8.** Let $A = (S, \{f\} \cup Y)$ be a one-point automaton with $n$ states and the transitive group $G = \langle Y \rangle$, then for every $e_i \in E_f$ there is a cyclic pair $(e_i, d_f)g_i$, $1 \leq i \leq k-1$, in the relation $\pi_{n-1}$.

**Proof.** Let $e_i \in E_f$, then consider the relation:

$$\rho_i = \{(e_i, d_f)g : \; g \in G,\; l_Y(g) \leq n-1\}.$$

It follows from formula (7) that $\pi_{n-1} = \rho_1 \cup \cdots \cup \rho_{k-1}$. Then consider the digraph $\Gamma_i = (S, \rho_i)$, which will be a subgraph of the digraph $\Gamma(\pi_{n-1})$. From the transitivity of $G$ it follows for every vertex $s \in S$ there is a permutation $g$ such that $(e_i)g = s$ and $l_Y(g) \leq n-1$. So from every vertex $s \in S$ of the digraph $\Gamma_i$ goes out at least one arc $(s, (d_f)g)$ and there are no loops in this digraph. So, we can walk along the arcs of $\Gamma_i$, starting from an arbitrary vertex, until we reach an already visited vertex, thus getting a cycle in the digraph $\Gamma_i$. Thus, the lemma is proved. □

**Theorem 9.** Let $A = (S, \{f\} \cup Y)$ be a one-point automaton with $n$ states, the transitive group $G = \langle Y \rangle$ and the strongly connected relation $\pi_A$, then:

$$\mathrm{msc}(A) \;\leq\; 2n-3.$$

**Proof.** Let us put $\rho_0 = C(\pi_{n-1})$, where the relation $\pi_{n-1}$ is defined by formula (7). Consider the cyclic digraph $\Gamma(\rho_0)$, then from Lemma 8, it follows that $\Gamma(\rho_0)$ contains a cycle, so it will be a disjoint union of at most $n-1$ strongly connected components. From Lemma 8, we also conclude that the relation $(\rho_0)G$ is strongly connected, since $\pi_0 \subseteq (\rho_0)G$ and therefore $(\pi_0)G \subseteq (\rho_0)G$. Thus, the following increasing chain of cyclic relations ends with a strongly connected relation:

$$\rho_0 \subset \rho_1 \subset \cdots \subset \rho_q , \tag{10}$$

where $\rho_i = (\rho_{i-1})Y \cup \rho_{i-1}$, $1 \le i \le q$, and $\rho_q^* = S \times S$.

So at each step in the chain (10) the number of strongly connected components of the digraphs $\Gamma(\rho_i)$ reduced at least by one, therefore $q \le n-2$. Since $\rho_0 \subseteq \pi_{n-1}$, then $\rho_q \subseteq \pi_{n-1+q} \subseteq \pi_{2n-3}$. Thus, the relation $\pi_{2n-3}$ is strongly connected and the theorem is proved. □

From the inequality (9) and Theorem 7, we have $\mathrm{at}(A) \le 2n-2$. From this and Lemma 3 we obtain the following upper bound for the reset threshold.

**Corollary 10.** Let $A = (S, \{f\} \cup Y)$ be a one-point automaton with $n$ states, the transitive group $G = \langle Y \rangle$ and the strongly connected relation $\pi_A$, then $\mathrm{rt}(A) \le 2(n-1)(n-2)+1$.

Thus, the question arises when the relation $\pi_A$ will be strongly connected. We have partial answers to this question. For directable automata with simple idempotents, the relation $\pi_A$ is strongly connected [13]. In [4] it was shown that, for an automaton with a full monoid of transformations the relation $\pi_A$ is also strongly connected. However, this result can be strengthened, as we shall see in the next paragraph.

## 7. Automata with a primitive group

Let $G \le \mathrm{Aut}_n(S)$ be a permutation group on the set $S$ and $(s,t)G$ be its orbital, then the digraph $(S,(s,t)G)$ is called the orbital digraph of $G$. A group $G$ on the set $S$ is called primitive if its congruences are only the diagonal $\Delta_S$ and the square $S \times S$. In the sequel, we need the following theorem [6].

**Theorem 11** (Higman-Sims). The transitive permutation group $G$ on the set $S$ is *primitive* if and only if every its orbital digraph is strongly connected.

As the following statement shows, for a primitive group the requirement of transitivity is redundant [5].

**Proposition 12.** Let $G$ be a primitive permutation group on the set $S$, then $G$ is transitive.

**Proof.** Recall that $|S| > 2$. Let $U = (s)G$ be an orbit of $G$, then $|U| > 1$, since in the other case the partition $\{U, S\setminus U\}$ will be a non-trivial proper congruence. So from the primitivity of $G$ follows that the Rees congruence $\sigma(U) = (U \times U) \cup \Delta_S$ will be universal $\sigma(U) = S \times S$. Thus, $U = S$ and the proposition is proved. □

**Theorem 13.** Let $A = (S, \{f\} \cup Y)$ be a one-point automaton with $n$ states and the primitive group $G = \langle Y \rangle$, then its relation $\pi_A$ is strongly connected and $\mathrm{rt}(A) \leq 2(n-1)(n-2)+1$.

**Proof.** According to Theorem 11 each orbital $(e_i, d_f)G$, $1 \leq i \leq k-1$, is strongly connected. Therefore, the relation $\pi_A$ is also strongly connected, since $(e_i, d_f)G \subseteq (\pi_0)G = \pi_A$. Thus, the automaton $A$ satisfies all conditions of Theorem 9, and then according to corollary 10 we obtain the upper bound of reset threshold $\mathrm{rt}(A) \leq 2(n-1)(n-2)+1$. □

The above theorem provides a partial answer to Problem 12.36 of [6]. From Theorems 7 and 13, we have the following statement.

**Corollary 14.** A weakly singular automaton with the primitive group is completely reachable and $\mathrm{rt}(A) \leq 2(n-1)(n-2)+1$.

Let $M$ be a monoid on the set $S$, then we denote by $\mathrm{sr}(M)$ the maximal rank of its singulars $\mathrm{sr}(M) = \max\{rk(f) : f \in M\setminus \mathrm{Aut}_n(S)\}$. The following result may be useful for transformation monoids.

**Lemma 15.** Let $M$ be a monoid on $S$, which contains a one-point map of rank $\mathrm{sr}(M)$. Then every generating set $X$ of $M$ contains a one-point map of rank $\mathrm{sr}(M)$.

**Proof.** Consider an automaton $A = (S, X)$ with the transition monoid $M = \langle X \rangle$. Let $f \in M$ be a one-point map of the rank $\mathrm{sr}(M)$, then $f$ is generated by a sequence of generators from $X$. Consider the first singular $f_1$ from this sequence; we show that $\mathrm{rk}(f) = \mathrm{rk}(f_1)$. If $f_1$ is not one-point, then any next applied generator either decreases the rank of $f_1$ or leaves it the same, giving a map of the same kernel type. Thus, it is not possible to generate $f$ using $f_1$.and therefore $f_1$ is one-point.□

**Theorem 16.** If a monoid $M$ on the set $S$ contains a one-point map of rank $\mathrm{sr}(M)$ and a primitive group of permutations $G$, then the following inequality holds:

$$\mathrm{rt}(M) \leq 2(n-1)(n-2)+1 . \quad (11)$$

**Proof.** Consider an automaton $A=(S,X)$, where $\langle X\rangle = M$, then there is a subset of generators $Y \subseteq X$ for the primitive group $G$. From Lemma 15 it also follows that there is at least a one-point singular $f \in X$ of rank $\mathrm{sr}(M)$. Consider the one-point automaton $B=(S,\{f\}\cup Y)$, then from Theorem 13 we have $\mathrm{rt}(B) \leq 2(n-1)(n-2)+1$. From this and inequality (3) it follows that $\mathrm{rt}(A) \leq \mathrm{rt}(B) \leq 2(n-1)(n-2)+1$. Since the inequality $\mathrm{rt}(A) \leq 2(n-1)(n-2)+1$ holds for every set of generators $X \subseteq M$, we obtain the inequality (11), and the theorem is proved. □

Note that the full monoid $\mathrm{End}_n(S)$ contains the primitive group $\mathrm{Aut}_n(S)$ and simple singulars of maximal possible rank $n-1$, therefore from Theorem 16 we have the following inequality [4] $\mathrm{rt}(\mathrm{End}_n(S)) \leq 2(n-1)(n-2)+1$.

In [4] was also obtained the lower quadratic bound for the reset threshold of the full monoid $n(n-1)/2 \leq \mathrm{rt}(\mathrm{End}_n(S))$. The proof was based on the reset threshold of the automaton $R_n$ described in [14], for which the equality $\mathrm{rt}(R_n) = n(n-1)/2$ was proved. However, this automaton is also interesting for its transition monoid.

Denote by $I_{n-1}(U)$ the symmetric inverse monoid (semigroup) on the set $U = \{1,\cdots,n-1\}$. This monoid consists of all partial bijections (one-to-one maps) of $U$ with the composition of bijections as binary relations. The empty bijection (relation) is the zero of this monoid. When a monoid contains zero its reset threshold is the maximum length of the zero. It is well known that $I_{n-1}(U)$ is an inverse semigroup, which may be defined as a regular semigroup with commuting idempotents [15].

**Theorem 17.** The equality holds $\mathrm{rt}\big(I_{n-1}(U)\big) = n(n-1)/2$ for all $n>1$.

**Proof.** When $n=2$, the theorem is trivial. Let $n>2$, then consider the automaton $R_n=(S,X)$ from [14], where $S=\{0,1,\cdots,n-1\}$ and $X = \{x_1,\cdots,x_{n-1}\}$. The maps from $X$ defined as follows $x_1=(0,0,2,\cdots,n-1)$, and $x_i = (i-1, i)$, $2 \leq i \leq n-1$, are transpositions. Then the map $x_1$ is a simple idempotent and other maps are permutations $Y = \{x_i : 2 \leq i \leq n-1\}$. Thus, $R_n = (S, \{x_1\} \cup Y)$ will be a weakly singular automaton with the group $G = \langle Y \rangle$, which isomorphic to the symmetric group $\mathrm{Aut}_{n-1}(U)$, where $U = S \backslash \{0\}$.

Let $M = \langle X \rangle$ and $f \in M$, then we put $T = S \backslash f^{-1}(0)$ and define the function $\varphi: M \to I_{n-1}(U)$ by the formula $\varphi(f) = f|_T$. The partial map $f|_T$ will be bijection $f|_T: T \leftrightarrow im(f) \backslash \{0\}$, since if $(s)f = (t)f$, where $s \neq t$, then $(s)f = (t)f = 0$. The following equality holds for all $f, g \in M$:

$$\varphi(f \circ g) = \varphi(f) \circ \varphi(g) \quad . \tag{12}$$

Indeed, if one side of the equality (12) is defined on a state $s \in S$ then the other is also defined and they are equal to $((s)f)g$. Thus, the function $\varphi$ is morphism (homomorphism) of monoids, because $\varphi(i_S) = i_U$.

Now suppose that $\varphi(f) = \varphi(g)$ for $f, g \in M$. If $(s)f = 0$, then $(s)g = 0$ and vice versa. If $(s)f \neq 0$ then $(s)g \neq 0$ and $(s)f(s) = (s)g$. So $f = g$ and the function $\varphi$ is an injection. Let $\alpha: \mathrm{dom}(\alpha) \leftrightarrow \mathrm{im}(\alpha)$ be an arbitrary partial bijection on $U$, then we can order $(s_1, \cdots, s_{n-1})$ the states in $U$ in such a way that elements from the subset $U \backslash \mathrm{dom}(\alpha)$ come first, and then come elements from the subset $\mathrm{dom}(\alpha)$. Since $G \cong \mathrm{Aut}_{n-1}(U)$ there is $g \in G$ such that $(i)g = s_i, 1 \leq i \leq n-1$. Let $U \backslash \mathrm{dom}(\alpha) = \{s_1, \cdots, s_k\}$, then we can direct all states from $U \backslash \mathrm{dom}(\alpha)$ to zero one by one by the map $f = f_1 \cdots f_k$, where $f_1 = x_1$, $f_i = x_i \cdots x_2 x_1$, $x_i \in Y$, $2 \leq i \leq k$. At last, there is a map $h \in G$ such that $(s)h = (s)\alpha$ for all $s \in \mathrm{dom}(\alpha)$, since $G$ is symmetric on $U$. Let us put $e = gfh$ and $T = \mathrm{dom}(\alpha)$, then $\varphi(e) = e|_T = \alpha$. Therefore the function $\varphi$ is a surjection and an injection.

Thus, the function $\varphi$ is the isomorphism of monoids $M$ and $I_{n-1}(U)$. From [14] and formula (4) we obtain the lower bound:

$$\frac{n(n-1)}{2} = \mathrm{rt}(R_n) \leq rt\big(I_{n-1}(U)\big) \quad .$$

The upper bound follows from [3], where was shown that any synchronizing automaton $A$ with $n$ states, which monoid has commuting idempotents, satisfies the inequality $\mathrm{rt}(A) \leq n(n-1)/2$. From this, we conclude that the inequality holds:

$$\mathrm{rt}\big(I_{n-1}(U)\big) \le \frac{n(n-1)}{2} ,$$

since in the monoid $I_{n-1}(U)$ idempotents commute. Thus, the theorem is proved. □

## 8. Conclusion

The concept of a monoidal semiautomaton makes it possible to operate with the transition monoid of an automaton and turns transformation monoids into basic objects on which the reset threshold function is defined. We have shown that in some cases, a non-trivial upper bound on the reset threshold can be established based purely on the properties of the monoid. Thus, prerequisites arise for more intensive use of the theory of semigroups in solving the Černý problem.

## 9. Acknowledgements

This work was supported in part by the National Science Centre, Poland under project number 2021/41/B/ST6/03691.